\documentclass[prd,aps,preprint,eqsecnum,floats,showpacs]{article}
\title{The heavy-quark pole masses in the Hamiltonian approach}
\author{A.M. Badalian and A.I. Veselov,\\ Institute of
Theoretical and Experimental Physics,\\ B.Cheremushkinskaya 25,
117218 Moscow,Russia \\ \\
B.L.G. Bakker, \\ Department of Physics,\
Vrije Universiteit,
Amsterdam, The Netherlands}
\date{}
\begin{document}
\maketitle

\newcommand{\be}{\begin{equation}}
\newcommand{\ee}{\end{equation}}

\def\la{\mathrel{\mathpalette\fun <}}
\def\ga{\mathrel{\mathpalette\fun >}}
\def\fun#1#2{\lower3.6pt\vbox{\baselineskip0pt\lineskip.9pt
\ialign{$\mathsurround=0pt#1\hfil##\hfil$\crcr#2\crcr\sim\crcr}}}

\begin{abstract}
From the fact that the nonperturbative self-energy contribution $C_{\rm
SE}$ to the heavy meson mass is small: $C_{\rm SE}(b\bar{b})=0$; $C_{\rm
SE}(c\bar{c})\cong -40$ MeV \cite{ref.01}, strong restrictions on the pole
masses $m_b$ and $m_c$ are obtained. The analysis of the $b\bar{b}$ and
the $c\bar{c}$ spectra with the use of relativistic (string) Hamiltonian
gives $m_b$(2-loop)$=4.78\pm 0.05$ GeV and $m_c$(2-loop)$=1.39 \pm
0.06$ GeV which correspond to the $\overline{\rm MS}$ running mass
$\bar{m}_b(\bar{m}_b)=4.19\pm 0.04$ GeV and $\bar{m}_c(\bar{m}_c)=1.10\pm
0.05$ GeV. The masses $\omega_c$ and $\omega_b$, which define the heavy
quarkonia spin structure, are shown to be by $\sim 200$ MeV larger than
the pole ones.

\end{abstract}

\section{Introduction}
\label{sec.I}

The spectrum of heavy quarkonia (HQ) is very rich and provides a unique
opportunity to study the static interaction in the infrared (IR)
region, and hyperfine and fine structure effects. To use that
opportunity one needs to know, besides such fundamental parameters as
the string tension and the strong coupling, also the heavy-quark mass,
which cannot  directly be measured since a quark is not observed as a
physical particle. Therefore the quark mass $m_Q$ has to be determined
indirectly, e.g.  from the study of hadronic properties like $e^+e^-\to
b\bar{b}$, hadronic decays, and the $Q\bar{Q}$ spectra.

In the QCD Lagrangian the mass parameter  depends on the
renormalization scheme and by convention this current mass is taken in
the $\overline{\rm MS}$  scheme. In perturbation theory it is
convenient to introduce the pole quark mass, i.e. the pole of the quark
propagator, and at present the $\overline{\rm MS}$ pole mass is known
to three-loops \cite{ref.02,ref.03}:
\be
m_Q=\bar{m}_Q(\bar{m}_Q)\left\{ 1+\frac{4}{3}
\frac{\alpha_s(\bar{m}_Q)}{\pi} +\xi_2\Biggl
(\frac{\alpha_s}{\pi}\Biggr )^2 +\xi_3\Biggl
(\frac{\alpha_s}{\pi}\Biggr )^3 \right \},
\label{eq.01}
\ee 
where the Lagrangian current masses,
\be
 \bar{m}_b(\bar{m}_b) = (4.25\pm 0.25)~{\rm GeV},\quad
 \bar{m}_c(\bar{m}_c)=(1.20 \pm 0.20)~{\rm GeV},
\label{eq.02}
\ee
are known now with an accuracy of 17\% (6\%) respectively for the $c$
quark ($b$ quark). Most calculations of the pole masses $m_b$ and $m_c$
have been done in the QCD sum rules approach \cite{ref.04} , lattice
QCD \cite{ref.05}, and different perturbative approaches
\cite{ref.02,ref.03}.

For three decades many properties of HQ like the spectra,
electromagnetic transitions, hadronic and semileptonic decays, were
successfully studied in different potential models (PM)
\cite{ref.06}-\cite{ref.13}, however, the heavy quark masses used in PM
are considered ``to make sense in the limited context of a particular
quark model'' \cite{ref.02} i.e. as fitting parameters.

However, in the last decade the situation has changed and in
Ref.~\cite{ref.11} the relativistic (string) Hamiltonian was
derived directly from the QCD Lagrangian, starting with the
gauge-invariant meson Green's function in Fock-Feynman-Schwinger
(FFS) representation.  In Refs.~\cite{ref.11}-\cite{ref.13} it was
established that for the orbital angular momentum $L\leq 5$ and not too large
string corrections, as in  HQ, the string Hamiltonian reduces to
the well-known Hamiltonian $H_0$ used in the relativized potential
model (RPM) for many years \cite{ref.07,ref.08}:
\be
 H_0 = 2\sqrt{\vec{p}^{\, 2}+m^2_q} + V_{\rm static}(r)
\label{eq.03}
\ee
It follows from the derivation of $H_0$ in Ref.~\cite{ref.11} that the
mass $m_q$ in (\ref{eq.03}) coincides with the $\overline{\rm MS}$ running
mass $\bar{m}_q(\bar{m}_q)$, if the perturbative interaction
is neglected, or with the pole mass $m_Q$ (\ref{eq.01}) if the perturbative
self-energy corrections are taken into account. Therefore this
Hamiltonian can be used to extract the pole mass $m_Q$ from the
analysis of the HQ spectrum.

Nevertheless, if one looks at the heavy-quark masses used in PM, a
large variety of $m_b$ and $m_c$ values can be found in different
analyses:  $m_c$ in the range 1.30 GeV $\div$ 1.84 GeV and  $m_b$ in
the range 4.20 $\div$ 5.17 GeV \cite{ref.07}-\cite{ref.10}. The main
reason behind this wide spread in the values of  $m_b$ and $m_c$ (even
for the same Hamiltonian $H_0$) is the presence  of a negative
arbitrary constant $C_0$ in the mass formula (or in the chosen static
potential). We give three examples: in Ref.~\cite{ref.07} $m_c=1.327$
GeV and $C_0=0$ is used by the Wisconsin Group; in Ref.~\cite{ref.08}
$m_c=1.628$ GeV and $C_0=-253$ MeV (in both cases the Hamiltonian
Eq.~(\ref{eq.03}) was used); in Ref.~\cite{ref.06} $m_c=1.84$ GeV and
$C_0\cong -800$ MeV are taken, i.e. the magnitude of $C_0$ is always
larger for larger heavy-quark mass.

The meaning of the constant $C_0$ was understood recently and in
Ref.~\cite{ref.01} it was shown that the negative contribution to
the meson mass comes from the nonperturbative (NP) interaction of the
color-magnetic moment of a quark (antiquark) with the background (vacuum)
field.  Moreover, this self-energy NP term $C_{\rm SE}$ was analytically
calculated with 10\% accuracy \cite{ref.01} (see Eq.~(\ref{eq.A.12})):
\be
 C_{\rm SE} (nL) =-\frac{4\sigma}{\pi \omega_q(nL)}\eta(m_q)
\label{eq.04}
\ee
for a quark and an antiquark with equal masses. In the expression
Eq.~(\ref{eq.04}) $m_q$ is the pole mass which determines the factor
$\eta(m_q)$ Eq.~(\ref{eq.A.11}) while
\be
 \omega_q(nL)= \left\langle \sqrt{\vec{p}^{\, 2}+m^2_q} \right\rangle_{nL}
\label{eq.05}
\ee
is the dynamical quark mass. For low-lying states in charmonium and
bottomonium $\omega_Q$ turns out to be $\sim 200$ MeV larger $m_Q$

The essential fact (for light and heavy-light mesons) is that $C_{\rm
SE}(nL) $ depends on the quantum numbers and just due to this the
correct intercept of the Regge-trajectory has been obtained in
Ref.~\cite{ref.13}. In HQ the situation appears to be much more
simple.  The factor $\eta(m_q)$ in Eq.~(\ref{eq.04}) depends on the
flavor through the pole mass $m_q$ and from the analytical expression
(\ref{eq.A.11}) one obtains a small value: $\eta_c\cong 0.35\div 0.27$
for $m_c$ in the range $1.37\div 1.70$ GeV and $\eta_b\cong 0.07$ for
$m_b \cong 4.7\div 5.0$ GeV. As a result 
$C_{\rm SE}(b\bar{b})\cong -3$ MeV, i.e. is compatible with zero,
and $C_{\rm SE}(c\bar{c})$ is also small:
\be
 C_{\rm SE}(b\bar{b})=0;~~~C_{\rm SE}(c\bar{c})=(-40\pm 10)~\mbox{\rm MeV}
\label{eq.06}
\ee

Thus the self-energy contributions to HQ states are well defined and
therefore  there is no opportunity anymore to vary the pole mass by
introducing a fitting constant. We shall show in this paper that  the
condition (\ref{eq.06}) puts strong restrictions on the values of
$m_b(m_c)$  needed to describe the $b\bar{b}(c\bar{c})$
spectrum. The extracted pole masses $m_b$ and $m_c$ in our analysis
will be determined with an accuracy better than 60 MeV and the main
uncertainty in their values comes not from the method used (for fixed
string tension and the strong coupling (or $\Lambda_{QCD})$ the
uncertainty is $\pm 10$ MeV) but from the uncertainty  in our knowledge
of the strong coupling in the IR region.  We shall show that HQ
spectra, in particular high excitations and the recently discovered 1D
state in bottomonium, can give very important information about the
strong coupling in the IR region.

Our analysis of HQ spectra shows that in bottomonium $m_b$(2-loop)
$< 4.70$ GeV and values $m_b >4.84$ GeV turn out to be incompatible
with the condition $C_{\rm SE}=0$. In charmonium the admittable $m_c$
values, $m_c=1.39\pm 0.06$ GeV, appear to be rather small and  agree with
the one obtained by Narison  with the use of the QCD sum rules for the
(pseudo)-scalar current \cite{ref.04}.  Our calculations of the HQ spectra
are done with the use of only three fundamental quantities: the string
tension, the QCD constant $\Lambda(n_f)$ and the pole mass $m_Q$. The
main emphasis in our fit lies on the excited (not ground) states.

This paper is organized as follows. In Section \ref{sec.II} the mass
formula, following from the relativistic Hamiltonian, as well as the
approximations to that, are presented and the notion of dynamical mass
is introduced. In Section \ref{sec.III} the static potential and the
strong coupling in the IR region which is defined in background
perturbation theory (BPT), are discussed. In Section \ref{sec.IV} from
the analysis  of the $b\bar{b}$ spectrum (with  special accent on
high excitations)  the pole mass $m_b$(2-loop) is obtained. In Section
\ref{sec.V} the pole mass $m_c$ is extracted from the $c\bar{c}$
spectrum. In Section \ref{sec.VI} our Conclusions are presented and in
the Appendix the method and NP self-energy term are discussed.

\section{The mass formula}
\label{sec.II}

The string corrections are small in  HQ and therefore the
simplified form of the Hamiltonian $H_R$ \cite{ref.11} may be used
(see Appendix):
\be
 H_R =\frac{\vec{p}^{\, 2}}{\omega} +\omega +\frac{m^2_q}{\omega}
 + V_{\rm static} .
\label{eq.07}
\ee
To derive this Hamiltonian in the FFS representation one needs to go over
from the proper time $\tau$ in the meson Green's function (\ref{eq.500})
to the actual time $t$ and at this point a new variable $\omega(t)$
must be introduced \cite{ref.11}:
\be
\omega(t) =\frac{1}{2}\frac{dt}{d\tau}
\label{eq.08}
\ee
This variable is the canonical coordinate and since $H_R$ does not
depend on its derivative, the requirement that the canonical momentum
$\pi_{\omega}=0$ is preserved in time, corresponds to the extremum
condition
\be
\dot{\pi}_{\omega} =\{\pi, H_R\}=\frac{\partial H}{\partial\omega}=0.
\label{eq.09}
\ee
From this extremum  condition the operator $\omega$,
\be
\omega =\sqrt{\vec{p}^{\, 2} + m^2_q},
\label{eq.10}
\ee
is defined as the kinetic energy operator. Substituting the definition
(\ref{eq.10}) into the $H_R$ (\ref{eq.07}) one arrives at the Hamiltonian
$H_0$ Eq.~(\ref{eq.03}):
\begin{equation}
 H_0=2\sqrt{{\bf p}^2+m^2_q} +V_{\rm static}(r),
\label{eq.0101}
\end{equation}
which
does explicitly not depend on the variable $\omega$. However, to
calculate different corrections to the meson mass, like spin and
string corrections, or the self-energy corrections which can be
considered as a perturbation and by derivation depend on the
$\omega$, we shall use the approximation when for a given state
the operator $\omega$ will be substituted by its average:
\be
\omega(nL) =\left\langle \sqrt{\vec{p}^{\, 2} + m^2_q} \right\rangle_{nL}
\label{eq.11}
\ee
This mass $\omega(nL)$ can be called the dynamical mass since its
difference with respect to the current (pole) mass $m_q$ is fully
determined by the dynamics. Note that for vanishing pole mass the
value of $\omega(nL)$ is finite and determines the constituent mass of
a light quark.

It is also important that perturbative corrections to the current mass,
which are essential at small quark-antiquark separations, $r\la 0.1$ fm
\cite{ref.15,ref.16}, are included in the pole mass $m_Q$ occurring in
$H_0$. On the other hand  the static potential $V_{\rm static}(r)$ is
well defined at $Q\bar{Q}$ separations $r\ga T_g\cong 0.2$ fm where
$T_g$ is the gluonic  correlation length \cite{ref.01}. The eigenvalues
of $H_0$, denoted as $M_0(nL)$,
\be
\left\{ 2\sqrt{\vec{p}^{\, 2} +m^2_Q}+ V_{\rm static}(r)\right \}
\psi_{nL}(r) = M_0(nL) \psi_{nL}(r)
\label{eq.12}
\ee
together with the self-energy term Eq.~(\ref{eq.06}) define the
heavy-meson masses. As shown in the Appendix, in bottomonium $C_{\rm
SE}=0$ and therefore the spin-averaged mass $M(nL)$ for a given
$b\bar{b}$ state coincides with the eigenvalue $M_0(nL)$:
\be
M(nL, b\bar{b}) = M_0(nL),
\label{eq.13}
\ee
while in charmonium from  Eq.~(\ref{eq.A.11}) $C_{\rm SE}\cong -40$ MeV and
\vspace{2mm}
 \be
M(nL,c\bar{c})=M_0(nL) +C_{\rm SE} .
\label{eq.14}
\ee
There exist two approximations to the solution of the spinless Salpeter
equation (SSE) (\ref{eq.12}) leading to two approximations to the meson
mass $M(nL)$. First, the nonrelativistic (NR) approximation when the mass
$M_{\rm NR}(nL)$ is given by:
\be
 M_{\rm NR}(nL)=2m_Q +E^{\rm NR}_{nL} (m_Q) + C_{\rm SE},
\label{eq.15}
\ee
where $E^{\rm NR}_{nL}(m_Q)$ is the eigenvalue of the Schr\"odinger
equation with the reduced mass equal to $\frac{1}{2}m_Q$.

There is also another, so called ``einbein'' approximation (EA) to
the solutions of SSE (\ref{eq.12}), where the meson mass given by
\be
 M_{\rm EA}(nL) =\omega(nL) +\frac{m^2_Q}
 {\omega_{nL}}+E_{nL}(\omega_Q) +C_{\rm SE}
\label{eq.16}
\ee
appears to be closer to the exact solution $M(nL)$ than in the NR
approximation \cite{ref.12}. In EA the binding energy
$E_{nL}(\omega_Q)$ is defined as the solution of the Schr\"odinger
equation with the reduced mass equal to $\frac{1}{2}\omega_Q(nL)$ (not
the pole mass $\frac{1}{2}m_Q$) while $\omega_Q(nL)$ is to  be defined
from the selfconsistent equation:
\begin{equation}
 \frac{\partial M_{\rm EA}}{\partial \omega} = 0, \; {\rm or}\;
 \omega_{nL} = \frac{m^2_Q}{\omega_{nL}}-\omega_{nL}
 \frac{\partial E_{nL}(\omega)}{\partial \omega_{nL}}.
\label{eq.161}
\end{equation}
Through $\omega_{nL}$ in EA the relativistic corrections are taken into
account in the mass formula (\ref{eq.16}). Moreover,  owing to the special
form of the mass formula (\ref{eq.16}) the choice of $\omega(b\bar{b})
\cong 5.0$ GeV turns out to be compatible with the condition $C_{\rm
SE}=0$, while in NR approximation the admittable values of $m_b$ are
about 200 MeV smaller.

It is worthwhile to notice that in bottomonium where both $\omega_Q(nL)$
and $m_Q$ are large,  around 5 GeV, the difference between NR, EA,
and relativistic cases is small, $| \delta_R | =M(nL) -M^{\rm NR}(nL)$
is about $10\div 20$ MeV. In charmonium the difference depends on the
quantum numbers and for high excitations can reach $\sim 100$ MeV (see
the discussion in Section \ref{sec.V}).

\section{Static potential}
\label{sec.III}

The static potential contains perturbative and NP contributions
where the NP linear potential can directly be derived from the meson
Green's function if the $q\bar{q}$ separation is larger than the
gluonic correlation length. From the analysis of the Regge
trajectories  of light and heavy-light mesons  the value of the
string tension, $\sigma=0.185 \pm 0.005$ GeV$^2$ \cite{ref.13}, is
fixed while the perturbative interaction in coordinate space can be
presentread in the form,
\be
 V_P(r) =-\frac{4}{3} \frac{\alpha_{\rm static}(r)}{r},
\label{eq.17}
\ee
where $\alpha_{\rm static}(r)$ is well known only in the
perturbative region, i.e.  at very small distances, $r \la 0.1$ fm
\cite{ref.15,ref.16}.

However,  the r.m.s. radii in bottomonium and charmonium, span a very
wide range:
\begin{eqnarray}
 & &  R(\Upsilon(1S))=0.2\mbox{~fm},\quad R(\chi_b(1 P))=0.4\mbox{~fm}, 
 \quad R(\chi_b(2P))=0.65\mbox{~fm},
 \nonumber \\
 & & R(\Upsilon(4S))=0.9\mbox{~fm}, \quad R(\Upsilon(6S))\cong 1.3\mbox{~fm},
\label{eq.18}
\end{eqnarray}
and
\begin{eqnarray}
 & & R(J/\psi)\cong 0.4\mbox{fm}, \quad R(\chi_c(1P))=0.6\mbox{fm}, \quad 
 R(\psi(1D)=0.8\mbox{fm} \nonumber \\
 & & R(\psi (3S))=1.1\mbox{fm}, \quad R(\psi(4S)\cong 1.40\mbox{fm}.
\label{eq.19}
\end{eqnarray}
Apparently,  with the exception of $\Upsilon(1S)$ the sizes of these
states lie outside the perturbative region.

Therefore the problem arises how to define the strong coupling
$\alpha_{\rm static}(r)$ at all distances, in particular in the IR
region. In PM it has always been assumed and later this fact has been
supported by direct measurement of the static potential in lattice QCD,
that the strong coupling freezes and reaches a critical (saturated)
value at large $r$. Unfortunately, at present there is no consensus
about the true value of $\alpha_{\rm crit}$ and  different values were
used. In the phenomenological analysis of Ref.~\cite{ref.08}) $\alpha_{\rm
crit}=0.60$ was used, but in analytical perturbation theory \cite{ref.17}
the large value $\alpha_{\rm crit}= 4\pi/\beta_0\cong 1.4$ appeared.
In the background perturbation theory (BPT) which will be used here,
$\alpha_{\rm crit}$ is smaller and fully defined by $\Lambda_{QCD}$
\cite{ref.16,ref.18}. For the definition of $\alpha_{\rm static}(r)$
it is better to start with the vector coupling in momentum space:
\be
 V_{\rm B}(q)=-4\pi C_F\frac{\alpha_{\rm B}(q)}{q^2}
\label{eq.20}
\ee
This background coupling $\alpha_{\rm B}(q)$ is defined in Euclidean momentum
space at all $q^2$, including $q^2=0$, i.e. it has no Landau singularity,
\be
 \alpha_{\rm B}(q,\mbox{2-loop})=\frac{4\pi}{\beta_0 t_{\rm B}}
 \left\{1-\frac{\beta_1}{\beta_0^2}\frac{\ln t_{\rm B}}{t_{\rm B}}\right\}.
\label{eq.21}
\ee
The logarithm
\be
 t_{\rm B} = \ln\frac{q^2+M^2_{\rm B}}{\Lambda^2_{\rm B}}
\label{eq.22}
\ee
contains the background mass $M_{\rm B}$ which appears due to the
interaction of a gluon with the background field at small $q^2$. This mass
$M_{\rm B}\cong 1$ GeV has the meaning of the lowest hybrid excitation:
$M_{\rm B}=M(Q\bar{Q}g)-M(Q\bar{Q})$ \cite{ref.19} and from the comparison
with the static potential $M_{\rm B}$ determined on the lattice, was
found to be equal to 1 GeV \cite{ref.16}.  The $t_{\rm B}$ (\ref{eq.22})
coincides in form with the parametrization of $\alpha_s(q)$ suggested
in Refs.\cite{ref.20} where instead of the background mass $M_{\rm B}$
two gluonic masses ($2m_g$) enter. However, the physical gluon  cannot
have a mass while $M_{\rm B}=M(Q\bar{Q}gg)-M(Q\bar{Q}g)$ is a well defined
physical quantity ($M_Q$ is supposed to be large) and can be calculated
in different theoretical approaches \cite{ref.20a} and on the lattice.

By definition $\alpha_{\rm B}(q)$ has the correct asymptotic freedom
(AF) behavior at large $q^2$ and in this region the connection between
the vector coupling $\alpha_{\rm B}(q)$ and $\alpha_s(q)$ in the
$\overline{\rm MS}$ renormalization  scheme is very simple, so that the
QCD constant $\Lambda_V$ (in the vector-scheme) can be expressed through
$\Lambda_{\overline{\rm MS}}$ \cite{ref.21}:
\be
 \Lambda^{(n_f)}_{\rm B} =\Lambda^{(n_f)}_{\overline{\rm MS}} \exp \left(
 \frac{a_1}{2\beta_0}\right)
\label{eq.23}
\ee
Here $a_1=\frac{31}{3}-\frac{10}{9}n_f$. At present the value
$\Lambda^{(5)}_{\overline{\rm MS}}$ (2-loop) $ = 215\pm 15$ MeV
is established from  high energy processes \cite{ref.02}, while in
quenched QCD the value $\Lambda^{(0)}_{\overline{\rm MS}}=240\pm 20$
MeV was calculated on the lattice \cite{ref.22}. Then from the relation
(\ref{eq.23}) it follows that in quenched approximation the QCD constant
in the vector scheme (\ref{eq.23}) has the value

\be
\Lambda^{(0)}_{\rm B} =385\pm 30~\mbox{MeV}
\label{eq.24}
\ee
and for $n_f=5$
\be
 \Lambda^{(5)}_{\rm B} =290\pm 30~\mbox{MeV} .
\label{eq.25}
\ee
Our choice of $\Lambda_{\rm B}$ in this paper will be in accord with
the numbers (\ref{eq.24}) and (\ref{eq.25}).

In coordinate space the background coupling $\alpha_{\rm B}(r)$ is
defined as the Fourier transform of $\alpha_{\rm B}(q)$:
\be
 \alpha_{\rm B}(r)=\frac{2}{\pi}\int\limits^{\infty}_0 dq
 \frac{\sin qr}{q} \alpha_{\rm B}(q),
\label{eq.26}
\ee
so that the perturbative part of the static potential is
\be
 V_{\rm B}(r)=-\frac{4}{3} \frac{\alpha_{\rm B}(r)}{r}
\label{eq.27}
\ee
and the static potential is the sum of $V_{\rm B}(r)$ and the NP linear
potential
 \be V_{\rm static}(r)=\sigma r-\frac{4}{3}
\frac{\alpha_{\rm B}(r)}{r}.
\label{eq.28}
\ee
In phenomenology the approximation where the coupling $\alpha_{\rm B}(r)$
is constant is often used. This approximation is valid, because at
distances $r \ga 0.4$ fm the coupling $\alpha_B(r)$ is already approaching
the freezing, or critical, value.  Also at large distances, $r \ga 1.2$
fm, the confining linear potential, due to $q\bar{q}$ pair creation,
is becoming more flat \cite{ref.22a,ref.23} and this effect can be important
not only for light mesons but also for high excitations in charmonium.

Note that the critical values of $\alpha_{\rm B}(q)$ and $\alpha_{\rm
B}(r)$ coincide,
\be
 \alpha_{\rm B}(q=0) =\alpha_{\rm B}(r\to \infty)=\alpha_{\rm crit},
\label{eq.29}
\ee
and their characteristic values in two-loop approximation  are
given below, $(\Lambda^{(n_f)}\equiv \Lambda^{(n_f)}_{\rm B}$)
\begin{eqnarray}
 \alpha^{(0)}_{\rm crit} (\Lambda^{(0)}=385~\mbox{MeV})&  =& 0.428
 \nonumber \\
 \alpha^{(3)}_{\rm crit} (\Lambda^{(3)}=370~\mbox{MeV}) & =& 0.510
 \nonumber \\
 \alpha^{(4)}_{\rm crit} (\Lambda^{(4)}=340~\mbox{Mev}) & =& 0.515 .
\label{eq.30}
\end{eqnarray}

Our calculations show that the bottomonium spectrum appears to be rather
sensitive to the AF behavior  while in charmonium the approximation
$\alpha_{\rm B}=const$ can be used with good accuracy. It is also
instructive to look at the effective coupling $\alpha_{\rm eff}$ for
different $c\bar{c}$ and $b\bar{b}$ states, which can be defined as
\be
 \left\langle \frac{\alpha_{\rm B}(r)}{r}\right\rangle_{nL} =
 \alpha_{\rm eff} (nL) \left\langle r^{-1}\right\rangle_{nL}
\label{eq.31}
\ee
and is dependent on the quantum numbers (see Table \ref{tab.01}) and
about 20\% is smaller than $\alpha_{\rm crit}$.

\begin{table}
\caption{\label{tab.01} The effective coupling $\alpha_{\rm eff}(nL)$
(\ref{eq.31}) for different $b\bar{b}$ and $c\bar{c}$ states for the
static potential (\ref{eq.28}) with $m_b=4.78$ GeV, $m_c=1.45$ GeV,
$\alpha_{\rm crit}=0.547$;$\sigma=0.185$ GeV$^2$, $\Lambda^{(4)}=360$ GeV
$(n_f=4)$.}
\begin{center}
\begin{tabular}{|l|llllll|}
 \hline
 State & ~~1S & ~~2S& ~~3S& ~~4S & ~~5S & ~~6S\\
 \hline
 $\alpha_{\rm eff}(nL,b\bar{b}$) & 0.386 & 0.419 &0.427 & 0.430 &
 0.431 & 0.432 \\$\alpha_{\rm eff}(nL,c\bar{c}$) & 0.441 & 0.447&
 0.446 & 0.445& ~~--& ~~-- \\
 \hline
 State & ~~1P & ~~2P& ~~3P& ~~1D & ~~2D & ~~1F\\ \hline
 $\alpha_{\rm eff}(nL,b\bar{b}$) & 0.474 & 0.480 &0.482 & 0.508 &
 0.508 & 0.520 \\$\alpha_{\rm eff}(nL,c\bar{c}$) & 0.510& 0.505&
 ~~-- & 0.531 & 0.526 & ~~-- \\
 \hline
\end{tabular}
\end{center}
\end{table}

From Table \ref{tab.01} one can see that in bottomonium due to the
different character of the wave functions the effective coupling  is
smaller for the $nS$ states and rather large for orbital excitations
like the $nD$ states.

\section{Bottomonium}
\label{sec.IV}

To extract the pole mass $m_b$ the $b\bar{b}$ spectrum will be studied
here as a whole. We mostly ignore the ground state -- the $\Upsilon(1S)$
mass, for which high perturbative corrections can be important
\cite{ref.03}, but rather concentrate on the following experimental
splittings \cite{ref.02,ref.14}
\begin{eqnarray}
 & &  M_{\rm cog}(1D) - M(1P) \cong M(1 ^3D_2)-M(1P) = 261.8 \pm 1.8
~\mbox{MeV}~\mbox{(exp)} \nonumber \\
 & & M(2P) - M(1P) = 360.0\pm 1.2 ~\mbox{MeV} ~\mbox{(exp)} .
\label{eq.32}
\end{eqnarray}
It is important that all $1P$, $2P$, and $1D$ states lie below the
$B\bar{B}$ threshold  and have  no  hadronic shifts. Also for these
splittings the small relativistic corrections are partly cancelled and
the calculations can be done either with the use of SSE or in NR
approximation.

First, we consider the case with $\alpha_{\rm static}=const$ and give the
$b\bar{b}$ spectrum  in Table \ref{tab.02} for three sets of parameters
with different $m_b$:
\begin{eqnarray}
 & {\rm  Set~I} & m_b=4.727~\mbox{GeV},~~~\sigma=0.20~\mbox{GeV}^2,~~~
\alpha_{\rm static}=0.3345, \nonumber \\
 & {\rm Set~II} & m_b=4.765~\mbox{GeV},~~~\sigma=0.19~\mbox{GeV}^2,~~~
 \alpha_{\rm static}=0.390, \nonumber \\
 & {\rm Set~III} & m_b=4.778~\mbox{GeV},~~~\sigma=0.188~\mbox{GeV}^2,~~\alpha_{\rm static}=0.415.
\label{eq.33}
\end{eqnarray}
In all cases $C_{\rm SE}=0$ is taken in the mass formula (\ref{eq.13}).

\begin{table}
\caption{\label{tab.02} The spin-averaged masses $M(nL)$ (in GeV) in
bottomonium for the potential (\ref{eq.28}) with the parameters
(\ref{eq.33}) (NR case).}
\begin{center}
\begin{tabular}{|l|llll|}
 \hline
 State & $Set~I$ & $Set~II$  & $Set~3$ & experiment\\
 \hline
 1S & 9.460 & 9.406$^{a)}$& 9.379$^{a)}$  & 9.4603$\pm$ 0.0003\\
 2S & 10.013 & 10.001${a)}$& 9.988  & 10.233$\pm$ 0.0003 \\
 3S & 10367 & 10.359 & 10.359 & 10.3552$\pm$0.0005 \\
 1P & 9.900 & 9.900 & 9.900 & 9.901$\pm $ 0.0006 \\
 2P & 10.267 & 10.267& 10.270 & 10.260$\pm$ 0.0006\\
 1D &10.150 & 10.156 & 10.161 & 10.1612$\pm$ 0.0016 \\
 \hline 
 \multicolumn{5}{|c|}{Above $B\bar{B}$ threshold, $M_{\rm th}$=10.558 GeV }\\
 \hline 
 4S & 10.659 & 10.647 & 10.649 & 10.5800$\pm $ 0.0035 \\
 5S  & 10.917 & 10.900 & 10.902 & 10.865$\pm$0.008\\
 6S & 11.146 & 11.131 & 11.132 & 10.019$\pm$0.008\\
 \hline 
\end{tabular}
\end{center}
$^{a)}$ The mass $M(1S)$ increases by an amount of $\sim 50\div 80$
MeV if the AF correction is taken into account.  
\end{table}

From the masses presented in Table \ref{tab.02} one can see that

\begin{enumerate}
\item  For small $m_b=4.727 $ GeV (Set I) the mass $M(1D)$ appears to
be about 10 MeV lower than the experimental value even for very large
$\sigma = 0.20$ GeV${}^2$.

\item For Set II and Set III almost identical fits are obtained with
exception of the 1D state where good agreement with experiment can
be reached only for a larger value of the coupling, as for Set III.

\item The spin-averaged $1D-1P$ splitting,
\be
 \Delta=M_{\rm  cog}(1D)-M_{\rm cog}(1P),
\label{eq.34}
\ee
has remarkable properties -- it is practically independent of the
relativistic correction $\delta_R$ and small variations of the string
tension, and therefore $\Delta$ can be considered the best and very
stable criterium to determine the critical value of the strong coupling
as well as the pole mass $m_b$.

\item In NR approximation Eq.~(\ref{eq.15}) and for SSE for the $b$-quark
masses $m_b \leq 4.70$ GeV or $m_b \geq 4.8$ GeV the condition
$C_{\rm SE}=0$ cannot be combined with a reasonably good fit to the
$b\bar{b}$ spectrum.

However, if one uses EA Eq.~(\ref{eq.16}) instead of the NR mass formula
Eq.~(\ref{eq.15}) the values of the dynamical mass $\omega_b(nL)$
are larger and the difference $\omega_b(nL)-m_b$ varies in the range
$(180\div 300$ MeV) (see Table 3), from which one can see that the
dynamical mass $\omega_b(nL)$ is slightly different for different $nL$
states. The $b\bar{b}$ spectrum calculated with $\omega_b(nL)=5.0$ GeV and
$m_b=4.78$ GeV with the use of the mass formula (\ref{eq.16}) gives values
coinciding within $\pm 5$ MeV with those from Table 2.  \end{enumerate}

\begin{table}
\caption{\label{tab.03} The dynamical mass $\omega_b(nL)$ (in GeV) and the
difference between the dynamical mass and the current quark mass (in MeV)
for SSE with Cornell potential $(m_b=4.78$ GeV, $\sigma=0.185$ GeV$^2$,
$\alpha_{\rm static}=0.4125$).}
\begin{center}
\begin{tabular}{|l|r|r|r|r|r|r|r|r|r|}\hline
 State & 1S & 2S& 3S& 4S& 5S& 6S & 1P&2P& 1D \\
 \hline
 $\omega(nL)$ & $5.043 $ & 5.008 & 5.028 & 5.057 & 5.088 & 5.127 &
 4.959 & 4.989 & 4.962 \\
 \hline
 $\omega(nL)-m_b$   &263 & 228 & 248 & 277 & 308 & 347 & 179 & 209 & 182 \\
 \hline
\end{tabular}
\end{center}
\end{table}

Thus from our fits, when the coupling is taken constant,
the extracted value of the pole mass
\be
 m_b=4.76\pm 0.02~\mbox{GeV}~~~(\alpha_{\rm static}=const)
\label{eq.36}
\ee
is obtained.

The picture does not change much if the AF behavior  of $\alpha_{\rm
B}(r)$ in two-loop approximation is taken into account.

However, in this case the admissible values of $m_b$ appear to be larger
by $\sim $50 MeV. The $b\bar{b}$ spectrum for $m_b\cong 4.82$ GeV and
$m_b=4.83$ GeV for the number of the flavors $n_f=4,5$  and also in
quenched approximation is presented in Table 4.

\begin{table}
\caption{\label{tab.04} The $b\bar{b}$ spectrum defined  by the mass
formula (\ref{eq.15}) for the spinless Salpeter equation.}
\begin{center}
\begin{tabular}{|l|l|l|l|l|}\hline
 & Set A & Set B & Set C & \\
 & $m_b=4.816$ & $m_b=4.83$ GeV & $ m_b=4.817$ &\\
 & $\sigma=0.185 $ GeV$^2$ & $\sigma=0.19$ GeV$^2$ &
 $\sigma=0.185$ GeV$^2$ & \\
 State & $\Lambda^{(5)}=360$ MeV & $\Lambda^{(4)}$ =390 MeV &
 $\Lambda^{(0)}$(1-loop)=365 MeV & experiment${}^a)$ \\
 \hline
 1S & ~9.471 & ~9.478 & ~9.470 & ~9.460\\
 2S & 10.025 & 10.032 & 10.023 & 10.023 \\
 3S & 10.376 & 10.386 & 10.375 & 10.355\\
 1P & ~9.900 & ~9.901 & ~9.900 & ~9.900 \\
 2P & 10.271 & 10.278 & 10.266 & 10.260 \\
 1D & 10.159 & 10.162 & 10.152 & 10.161 \\
 \hline
\end{tabular} 
\\
${}^{a})$ The experimental errors of the masses are given in Tab.~\ref{tab.02}
\end{center}
\end{table}

As seen from Table \ref{tab.04} good agreement for Sets A and B is
obtained for the $1P$ and $1D$ states but for the $1S$ level the mass
is $\sim 10$ MeV higher than the $M(\Upsilon(1S))$ value. This fact can
be connected with the contribution of the 3-loop perturbative correction
which is neglected here.

Our conclusion is that for the vector constant $\Lambda^{(4)}=390$
MeV, or $\Lambda^{(5)} \leq 365$ MeV, and $\sigma=0.186 \pm 0.004$
GeV$^4$, the extracted pole mass lie in the narrow range
\be
 m_b (2-\mbox{loop})= 4.81\pm 0.02\;\mbox{GeV}.
\label{eq.361}
\ee
Thus, if the AF behavior of $\alpha_{\rm B}(r)$ is taken into account
the extracted pole mass is about 50 MeV larger than $m_b$ (\ref{eq.34})
for $\alpha_{\rm static}=const$.

Then combining Eqs.~(\ref{eq.36}) and (\ref{eq.34}) for different choices of
$\alpha_{\rm B}$ one obtains that the extracted pole mass of the $b$ quark
lies in the range
\be
 m_b=4.78\pm 0.05 ~\mbox{GeV}.
\label{eq.37}
\ee
Then by definition of the two-loop pole mass (\ref{eq.01}) $(n_f=5)$
where in the relation (\ref{eq.01}) the parameter $\xi_2$ is 
\begin{equation}
 \xi_2(n_f=5) =\left \{ -1.0414  \sum^{N_L}_k\left(1-\frac{4}{3}
 \frac{\bar{m}_{Q_k}}{\bar{m}_Q}\right) +13.4434 \right \},
\label{eq.371}
\end{equation}
and the sum over $k$ extends over the $N_L$ flavors $Q_k$
which are lighter than $Q$, one finds $\xi_2(n_f=5)\cong 9.6\div
9.7$, and for $\alpha_s(\bar{m}_b)=0.217$  it follows from
(\ref{eq.01}) that
\be
 \bar{m}_b(\bar{m}_b) = (4.19\pm 0.04)\,\mbox{GeV},
\label{eq.38}
\ee
This number for the $\bar{\rm MS}$ mass (\ref{eq.38}) appears to be
in good agreement with the conventional value from Ref.~\cite{ref.02}
but has smaller theoretical error, $\sim 50$ MeV, than the number quoted
in Eq.~(\ref{eq.02}).

\section{Charmonium}
\label{sec.V}

The $c\bar{c}$ spectrum has several differences in comparison
to bottomonium.

First, the self-energy contribution to the mass $M(nL)$ is nonzero,
about  $-40$ MeV, being practically the same for different $nL$ states,
and therefore can be taken constant for all states with the accuracy
$1\div 3$ MeV.

Secondly, relativistic corrections are not small in charmonium
and  $M(nL)$ is in the relativisitc case (R) always smaller, so
\be
 \delta_R(nL) = M(nL) - M^{\rm NR}(nL)
\label{eq.39}
\ee
is  negative. Note that the self-energy term must be the same in
both cases . However, if the spin-averaged mass of the 1S state,
$M(1S)$=3067 MeV, is used for the fit, then $C_{\rm SE}$ are different
in R and NR cases and $\delta_R(nL)$ has irregular behavior (see
Table \ref{tab.05}). From Table \ref{tab.05}, where $C_{\rm SE} =-35$
MeV in the R case and $C^{NR}_{\rm SE} =-57$ MeV in the NR case, one
can see that $\delta_R(nL)$ is \underline{positive} $(\sim 10\div 20$
MeV) for the $1P$ and $1D$ states; equal to zero for the $1F$ state,
and \underline{negative} for the $nS$ states, $2P$ and $2D$, and higher
states. It is important that for the $4S$ ($3S$) state $|\delta(nL)|$
is large, $\sim 100$ MeV (60 MeV) and therefore the $c\bar{c}$ spectrum
has to be calculated with a relativistic Hamiltonian.

\begin{table}
\caption{\label{tab.05} The $c\bar{c}$ spin-averaged masses $M(nL)$
(in MeV) in R and NR cases with the same static potential
($\alpha_{\rm static}=0.42$; $\sigma=0.18$ GeV$^2$), $m_c=1.41$~GeV
and different $C_{\rm SE}$ (MeV). The quantity $\delta_R$ is
the relativisitic shift $M(nL)-M^{\rm NR}(nL)$.}
\begin{center}
\begin{tabular}{|l|l|l|l|r|}
\hline
State & Experiment & R case & NR case & $\delta_R$\\
 & & $C^{\rm R}_{\rm SE}= -35$ &$C_{\rm SE}^{\rm NR}= -57$ & \\
 \hline
 1S & 3067$\pm$ 0.7 & 3067 & 3067 (fit) & 0\\
 2S & $3673\pm$ 6 & 3661 & 3688 & $- 27$ \\
 1P & 3525$\pm$ 0.6 & 3528 & 3510 & 18 \\
\hline
&
\multicolumn{4}{|c|}{Above $D\bar{D}$ threshold}\\
\hline
 1D & 3770$\pm$ 2.5 & 3823 & 3812 & 12\\
 & 3872$\pm$1.2 &&& \\
 2P & ~~~~-- & 3965 & 3984 & $-19$ \\
 1F & ~~~~-- & 4067 & 4067 & 0\\
 3S & 4040$\pm$10 & 4082 & 4141 & $-59$  \\
 2D & 4159$\pm$20 & 4200 & 4227 & $-27$  \\ 
 4S & 4415$\pm$6 & 4433 & 4527 & $-94$  \\
 \hline
\end{tabular}
\end{center}
\end{table}

The interesting observation is that while the ground state mass is fitted
well, the relativistic corrections to $M_{\rm NR}(nL)$ for the $1P$
and $1D$ states turn out to be \underline{positive},  (since a negative
value for $C^{\rm NR}_{\rm SE}$ in NR with larger magnitude was taken).

A third difference refers to the choice of $\alpha_{\rm B}(r)$. Since
the $c\bar{c}$ states have larger sizes than the $b\bar{b}$ ones  the
AF behavior of $\alpha_{\rm B}(r)$ appears to be less important in
charmonium and the approximation $\alpha_{\rm B}(r)$=constant is valid
with good accuracy. For example for two  sets of parameters:
\be
\begin{array}{lllll}
 {\rm Set~A}& m_c=1.42~{\rm GeV}, &\sigma=0.18~\rm{GeV},
 &\alpha_{\rm B}=0.42; & C_{\rm SE} =-35~{\rm MeV} \\
 {\rm Set~B} &m_c=1.42~{\rm GeV}, & \sigma=0.185~{\rm GeV},&
 \Lambda^{(4)}=360~{\rm MeV}, & C_{\rm SE}=-30~{\rm MeV}
\end{array}
\label{eq.40}
\ee
close values of $M(nL)$ in the relativistic case are obtained 
(see Table \ref{tab.06}).

\begin{table}
\caption{\label{tab.06} The spin-averaged masses in charmonium in $R$ case
for the static potential with the parameters given in Eq.~(\ref{eq.40}).}
\begin{center}
\begin{tabular}{|l|l|l|l|}\hline
State & Set A & Set B & Experiment\\ 
\hline
 1S & 3067 & 3067 & 3067$\pm$ 0.7\\
 2S & 3660 & 3668& 3673$\pm$ 8 \\
 1P & 3528 & 3510& 3525$\pm$ 0.6 \\
\hline
 \multicolumn{4}{|c|}{\underline{Above $D\bar{D}$ threshold }} \\ 
 \hline
 1D & 3823 &3805 & 3871.8$\pm$ 1.2 \\
 &&& 3770.$\pm$ 2.5\\
 2D & 4199 &4198 & 4159 $\pm$ 20\\
 3S & 4080 & 4109& 4040$\pm$ 10\\
 4S & 4424 &4459 & 4415$\pm$ 6\\
 2P & 3964 &3954& ~~~--\\
 \hline
\end{tabular}
\end{center}
\end{table}

As seen from Table \ref{tab.06} the higher excitations, like the 3S, 4S,
and 2D states, lie $\sim 40$ MeV higher than the experimental values. All
these states have large r.m.s. radii: $R(3S)=1.1$ fm, $R(4S)\cong 1.4$
fm, and $R(2D)\cong 1.4$ fm. At such distances  the confining potential
is flattening due to quark-antiquark pair creation \cite{ref.22a} and it
results in a correlated shift of the radial excitations down as it takes
place for light mesons \cite{ref.23}. This phenomenon can be illustrated
taking instead of the linear potential $\sigma_0 r$ the modified confining
potential $\sigma(r)~r$ which was proposed in Ref.~\cite{ref.23},
\be
 \sigma(r)=\sigma_0(1-\gamma_0f(r))~\mbox{with}~
 f(r) =\frac{\exp(\sqrt{\sigma_0}\;(r-a))}
 {B+\exp(\sqrt{\sigma_0}\;(r-a))}
\label{eq.41}
\ee
with the
parameters
\be
\sigma=0.185~\mbox{GeV}^2,~~~\gamma_0=0.40,~~~a=6.0~\mbox{GeV}^{-1},~~~
B=20 .
\label{eq.42}
\ee
For this set of parameters the $c\bar{c}$ spectrum (R case) is given
in Table \ref{tab.07} together with the one for the standard linear
potential $\sigma_0r$ with $\sigma_0=0.185$ GeV$^2$. The value $\alpha_{\rm
static}=0.42$ is taken in both cases.

\begin{table}
\caption{\label{tab.07} The comparison of the spin-averaged masses
in charmonium ($R$ case) for confining $\sigma_0r$ potential and
modified potential (\ref{eq.41}) ($m_c=1.42$ GeV, $C_{\rm SE}=-42$
MeV, $\alpha_{\rm static}=0.42$ in both cases).} 
\begin{center}
\begin{tabular}{|l|l|l|l|}
\hline
 State & $\sigma_0=const=$ & $\sigma=\sigma(r)$ with & experiment\\
 & $=0.185$ GeV$^2$ & parameters Eq.~(\ref{eq.42})&  \\
 \hline 
 1S &3068 & 3067 & 3067 \\
 2S & 3670& 3664 & 3672 \\
 3S & 4097 & 4077 & 4040$\pm$10 \\
 4S & 4454 & 4403 &4415$\pm$6 \\
 1P & 3535 & 3530 &3525$\pm$0.6 \\
 2P & 3979 & 3965 & ~~~-- \\
 1D & 3835 &  3828 & 3779$\pm$25 \\
 2D & 4217 & 4194 & 4159$\pm$ 20 \\
 \hline 
\end{tabular} 
\end{center} 
\end{table}
From Table \ref{tab.07} one can see that for the modified potential
$\sigma(r)\, r$ the masses $M(4S)$ and $M(3S)$ of the radial excitations
are shifted down by 50 MeV and 20 MeV respectively, and turn out to be
closer to the experimental values. It is also worthwhile to look at the
dynamical masses for the low-lying states which are larger than the pole
mass by the constant amount
\be
\omega_c(nL)-m_c \cong 220 \div 250~\mbox{MeV},
\label{eq.43}
\ee
while for high excitations this difference  can reach $300\div 340$ MeV
(see Table \ref{tab.08}).
\begin{table}
\caption{\label{tab.08} The dynamical masses $\omega_c(nL)$ for the
potential with the parameters Eq.~(\ref{eq.40}) and $m_c=1.42$ GeV (Set A).}
\begin{center}
\begin{tabular}{|l|l|l|l|l|l|l|l|l|}
 \hline
 State & 1S  & 2S  & 3S & 4S& 1P & 2P & 1D & 2D\\
 \hline
 $\omega_c(nL)$ in GeV & 1.65 & 1.69 & 1.74 & 1.76 & 1.63 &  1.69 & 1.66 &
 1.77 \\
 \hline
\end{tabular}
\end{center}
\end{table}
The observed difference between the dynamical and the pole mass can
be essential for such physical characteristics as the hyperfine and
fine-structure splittings, which are determined by the dynamical mass
\cite{ref.24} and it causes a decrease of the hyperfine splitting,
e.g. for the $2S$ state in charmonium~\cite{ref.25a}.

In our analysis the best fit to the $c\bar{c}$ spectrum together  with
the correct choice of the self-energy contribution Eq.~(\ref{eq.06})
gives the pole mass $m_c$ in the range
\be
 m_c(\mbox{2-loop})=1.39\pm 0.01 \mbox{GeV~(theory)}\pm 0.04(\alpha_{\rm B}) .
\label{eq.44}
\ee
Then the $\overline{\rm MS}$ running mass Eq.~(\ref{eq.02}) $(n_f=4$, the
coefficient $\xi_2\cong $ 10.5) from Eq.~(\ref{eq.44}) is
\be
\bar{m}_c(\bar{m}_c)=(1.10\pm 0.05)
\mbox{GeV}.
\label{eq.45}
\ee
The value obtained turns out to be in good agreement with the conventional
value for $\bar{m}_c(\bar{m}_c)$ \cite{ref.02}, but has a smaller
theoretical error.

\section{Conclusion}
\label{sec.VI}

Our study of the $b\bar{b}$ and $c\bar{c}$ spectra has been performed with
the use of the relativistic Hamiltonian $H_0$ and correct NP self-energy
contribution to the meson mass.

By derivation the kinetic part of $H_0$ contains the pole quark mass $m_Q$
and it can directly be extracted from the analysis of the $Q\bar{Q}$
spectrum. In our study all meson masses are expressed through only two
parameters: the string tension and the QCD constant $\Lambda^{(n_f)}$
(in the Vector scheme). In charmonium the strong coupling $\alpha_{\rm
B}(r)$  can be approximated by a constant with good accuracy.

The spin-averaged splittings like 1D-1P and 2P-1P in bottomonium and
2S-1P and 1P-1S in charmonium appear to be very sensitive to the chosen
freezing (critical) value of the strong coupling. A good description of
the HQ spectra was reached only if $\alpha_{\rm crit}$ was taken rather
large, $\alpha_{\rm crit}\cong 0.55\pm 0.02$ while the constant value
for $\alpha_{\rm eff}$ taken in the Coulomb potential is about 20\%
smaller. \\

From our analysis one can conclude that\\

(i) The dynamical quark mass $\omega_b(\omega_c)$ is about 200 MeV larger
than the pole mass $m_b(m_c)$ for low-lying states. This difference
should be taken into account when the spin structure in heavy quarkonia
is studied, and it is especially important in charmonium.\\

(ii) The pole masses, $m_b(2$-loop)=$4.78\pm 0.05$ GeV and $m_c=1.39\pm
0.06$ GeV, were extracted from our fit to the $Q\bar{Q}$ spectra which
correspond to the $\bar{\rm MS}$ running masses: $\bar{m}_b(\bar{m}_b)
=4.19\pm 0.04$ GeV and $\bar{m}_c(\bar{m}_c) =1.10\pm 0.05$ GeV. The
numbers obtained are in good agreement with the conventional values but
have smaller theoretical error. The error we found is small, because in
our analysis only one parameter, $\Lambda$ (or $\alpha_{\rm crit}$),
is actually varied while a second parameter--the string tension--was
taken the same as for light mesons.

\appendix

\section{Relativistic Hamiltonian}

Here we present the principal steps to derive the Hamiltonian $H_R$
Eq.~(\ref{eq.07}) and the NP self-energy term Eq.~(\ref{eq.04}) taken from
Refs.  \cite{ref.01,ref.11}. The starting point is the gauge-invariant
meson Green's function written in FFS representation \cite{ref.01,ref.24}
with the use of the QCD action
\be
 G_M (x,y) =\langle Tr~\Gamma_1~G_q(x,y) \Gamma_2~G_{\bar{q}}(x,y)
 \rangle_{\rm B}
\label{eq.A.01}
\ee
In Eq.~(\ref{eq.A.01}) the averaging goes over the background field
$B_{\mu}$ and $G_q(x,y)(G_{\bar{q}}(x,y))$ is the Euclidean quark
(antiquark) Green's function
\begin{eqnarray}
\label{eq.500}
 G_q(x,y) & = & (\bar{m}_q+\hat{D})^{-1}_{x,y} = (\bar{m}_q-\hat{D})_x
 (\bar{m}^2_q-\hat{D}^2)^{-1}_{x,y} \\ \nonumber
 & = & (\bar{m}_q-\hat{D})_x\int\limits^{\infty}_0 ds (Dz)_{xy}
 e^{-K} {\cal{R}}_a{\cal{R}}_{\rm B}{\cal{R}}_F,
\end{eqnarray}
where the factors
${\cal{R}}_a$, ${\cal{R}}_{\rm B}$, and ${\cal{R}}_F$ are given by,
\begin{eqnarray}
 {\cal{R}}_a & =& P_a \exp\left(ig\int\limits^x_y a_{\mu} dz_{\mu} \right),
 \nonumber \\
 {\cal{R}}_{\rm B}& =&
 P_{\rm B} \exp\left(ig\int\limits^x_y B_{\mu}dz_{\mu} \right),
 \nonumber \\
 {\cal{R}}_F & =& P_F~\exp\Biggl (\int\limits^s_0 g\sigma_{\mu\nu}
 F_{\mu\nu} d\tau\Biggr )
\label{eq.A.05}
\end{eqnarray}
(The factors corresponding to the antiquarks are defined similarly.)
Here $P_a$, $P_{\rm B}$, and $P_F$ are the ordering
operators of the matrices $a_{\mu},B_{\mu}$, and $F_{\mu\nu}$
respectively, and
\be
 \sigma_{\mu\nu} F_{\mu\nu}= \Biggl ( \begin{array}{ll}
 \vec{\sigma}\vec{H} & ~~\vec{\sigma} \vec{E} \\
 \vec{\sigma}\vec{E} & ~~\vec{\sigma}\vec{H} \end{array}\Biggr ) ,
\label{eq.A.06}
\ee
$\sigma_{\mu\nu} =(1/4i)(\gamma_{\mu}\gamma_{\nu}-\gamma_{\nu}\gamma_{\mu})$
represents the interaction of the quark (antiquark) color magnetic
moment with the NP field strength $F_{\mu\nu}$.

In Eq.~(\ref{eq.500}) the kinetic energy term $K$ is defined as the  integral
over the proper time $\tau$:
\be
 K=\overline{m}^2_q \, s +\frac{1}{4} \int\limits^s_0(\dot{z}_{\mu})^2 d\tau.
\label{eq.A.07}
\ee
The quark moving along the trajectory $z_{\mu}(\tau)$ interacts  with
the field of the valence gluon $a_{\mu}$ and by its color charge with
the NP background field $B_{\mu}$.

In Eq.~(\ref{eq.A.07}) the quantity $\overline{m}_q$ is the Lagrangian current
mass usually taken in the $\bar{\rm MS}$ renormalization scheme. The
factor ${\cal{R}}_a$ is responsible for the standard perturbative
corrections to the quark mass $\bar{m}_q$ (as in Eq.(\ref{eq.01})),
i.e. for the appearance of the pole mass in the QCD Action (Hamiltonian)
\cite{ref.03}. Finally, the factors ${\cal{R}}_a$ and $\bar{\cal{R}}_a$
(from the quark and the antiquark) provide the perturbative static
interaction \cite{ref.24}.

The other two factors ${\cal{R}}_{\rm B}$ and $\bar{\cal{R}}_{\rm B}$
(from the quark and the antiquark) in $G_M(x,y)$ Eq.~(\ref{eq.500})
are responsible for the full NP (string) dynamics and were considered
in detail in Ref.~\cite{ref.11}, where after several steps the meson
Green's function was presented in following form
\be
 G_M(r) =\int d\omega ~d\nu ~dr~\exp(-A_R),
\label{eq.A.08}
\ee
where the action $A_R$ in coordinate space  is expressed through
two auxiliary fields $\omega$ and $\nu$. Since this action (see
Ref.~\cite{ref.11}) does not depend on the derivatives $\dot{\omega}$
and $\dot{\nu}$, the integration over $\omega,\nu$ in Eq.~(\ref{eq.A.08})
is equivalent to the canonical quantization of the Hamiltonian $H_R$ which
corresponds to the action $A_R$. It results in the following Hamiltonian,
\begin{eqnarray}
 H_R & = & \frac{p^2_r +m^2_q}{\omega} +\omega(t)
 +\frac{\vec{L}^{\, 2}}{r^2} \Biggl [ \omega +2 \int^1_0
 d\beta\; \beta\, \nu(\beta)\Biggr ]^{-1}
 \nonumber \\
 & & +\frac{1}{2} \sigma^2r^2 \int\limits^1_0
 \frac{d\beta}{\omega(\beta)} +\frac{1}{2} \int\limits^1_0 d\beta\,\nu(\beta),
\label{eq.A.09}
\end{eqnarray}
where the field operator $\omega(t)$ is defined as
$\omega(t)=\frac{1}{2}\frac{dt}{d\tau}$ and $t$ is the actual time. With
the use of the extremal conditions $(\nu_0=\sigma r)$ and considering
the string corrections as a perturbation (the procedure is described
in Ref.~\cite{ref.13}) one obtains the Hamiltonian $H_0$. The terms
${\cal{R}}_F$ and $\bar{\cal{R}}_F$ (from quark and antiquark) provide
the NP self-energy contribution $C_{\rm SE}$ (gauge-invariant) to the
meson mass \cite{ref.01}, where for the quark (antiquark) the self-energy
correction $\Delta m^2_q$ to the pole mass $m_q$ appears to be expressed
through the string tension $\sigma$ and the factor $\eta$:
\be
 \Delta m^2_q (m_q)=-\frac{4\sigma}{\pi} \eta (m_q)
\label{eq.A.10}
\ee
The factor $\eta(m_q)$ was calculated in analytical form in
Ref.~\cite{ref.01} and for $m_q > T_g$, where $T_g$ is the gluonic
correlation length ($\delta=T_g^{-1})$, $\eta(m_q)$ is given by the
expression
\be
 \eta(m_q)=-\frac{3m^2_q\delta^3}{(m^2_q-\delta^2)^{5/2}} ~\arctan
 ~\frac{\sqrt{m^2_q-\delta^2}}{\delta}+\frac{\delta^2(2m^2_q
 +\delta^2)}{(m^2_q-\delta^2)^2} 
\label{eq.A.11}
\ee
A straightforward calculation gives $\eta(m_{\rm B}\cong 5.0)\cong 0.07$;
$\eta(m_q=1.70)=0.24$ and $\eta(m_q\cong 1.40)\cong 0.30$. Then the
unperturbed part of the string Hamiltonian Eq.~(\ref{eq.07}) acquires
the correction Eq.~(\ref{eq.A.10})
\begin{equation}
 H'_0 \to \frac{\vec{p}^{\, 2}  + m^2_q +\Delta m^2_q}{\hat{\omega}_q}
 + \hat{\omega}_q+V_{\rm static}=H_0+C_{\rm SE}
\label{eq.A.111}
\end{equation} 
with the self-energy correction
\be
 C_{\rm SE} =\frac{\Delta m^2_q}{\hat{\omega}_q}
 = -\frac{4\sigma}{\pi \hat{\omega}_q} \eta(m_q).
\label{eq.A.12}
\ee
If this self-energy correction is considered as a perturbation, the
operator $\hat{\omega}$ in Eq.~(\ref{eq.A.12}) can be replaced by the
average of this operator Eq.~(\ref{eq.05}), i.e. by the dynamical mass
\begin{equation}
 \hat{\omega}_q \to \omega_q =\langle \sqrt{\vec{p}^{\, 2}+m^2_q} \rangle_{nL}
\label{eq.A.121}
\end{equation}
and through $\omega_q$ the NP self-energy term $C_{\rm SE}$ appears to be
dependent on the quantum numbers $nL$. However, in bottomonium
\begin{equation}
 C_{\rm SE}(b\bar{b}) \cong -3~\mbox{MeV}
\label{eq.A.122}
\end{equation}
is small and can be neglected in the mass formula. In charmonium however,
for $m_c\cong 1.40$ GeV the factor $\eta_c=0.29$ and the value of
$C_{\rm SE}\cong-40$ MeV is obtained which is practically the same for
different $nL$ states because of the weak dependence of $\omega_c(nL)$
on the quantum numbers (see Table \ref{tab.08}).

\end{document}